\documentclass[12pt]{article}
\usepackage{graphicx}
\usepackage{epsfig}
\usepackage{epstopdf}
\DeclareGraphicsExtensions{.pdf,.eps,.png,.jpg,.mps}

\def\lsim{\mathrel{\rlap{\lower3pt\hbox{\hskip0pt$\sim$}}
     \raise1pt\hbox{$<$}}}         
\def\gsim{\mathrel{\rlap{\lower4pt\hbox{\hskip1pt$\sim$}}
     \raise1pt\hbox{$>$}}}         

\usepackage{amsmath}
\usepackage{amsfonts}

\begin{document}
\begin{titlepage}

\centerline{\Large \bf FX Options in Target Zone}
\medskip

\centerline{Peter P. Carr$^\sharp$\footnote{\, Professor Peter P. Carr, Ph.D., is the Chair of the Finance and Risk Engineering Department at NYU Tandon School of Engineering. Email: \tt pcarr@nyc.rr.com} and Zura Kakushadze$^\S$$^\dag$\footnote{\, Zura Kakushadze, Ph.D., is the President of Quantigic$^\circledR$ Solutions LLC,
and a Full Professor at Free University of Tbilisi. Email: \tt zura@quantigic.com}}
\bigskip

\centerline{\em $^\sharp$ Department of Finance and Risk Engineering, NYU Tandon School of Engineering}
\centerline{\em 12 MetroTech Center, Brooklyn, NY 11201}
\medskip
\centerline{\em $^\S$ Quantigic$^\circledR$ Solutions LLC}
\centerline{\em 1127 High Ridge Road \#135, Stamford, CT 06905\,\,\footnote{\, DISCLAIMER: This address is used by the corresponding author for no
purpose other than to indicate his professional affiliation as is customary in
publications. In particular, the contents of this paper
are not intended as an investment, legal, tax or any other such advice,
and in no way represent views of Quantigic$^\circledR$ Solutions LLC,
the website \underline{www.quantigic.com} or any of their other affiliates.
}}
\medskip
\centerline{\em $^\dag$ Business School \& School of Physics, Free University of Tbilisi}
\centerline{\em 240, David Agmashenebeli Alley, Tbilisi, 0159, Georgia}
\medskip
\centerline{(December 1, 2015; revised June 21, 2016)}

\bigskip
\medskip

\begin{abstract}
{}In this note we discuss -- in what is intended to be a pedagogical fashion -- FX option pricing in target zones with attainable boundaries. The boundaries must be reflecting. The no-arbitrage requirement implies that the differential (foreign minus domestic) short-rate is not deterministic. When the band is narrow, we can pick the functional form of the FX rate process based on computational convenience. With a thoughtful choice, the FX option pricing problem can be solved analytically. The European option prices are expressed via (fast converging) series of elementary functions. We discuss the general approach to solving the pricing PDE and explicit examples, including analytically tractable models with (non-Ornstein-Uhlenbeck) mean-reversion.
\end{abstract}
\medskip
\end{titlepage}

\newpage

\section{Introduction}

{}Foreign exchange (FX) rates in target zones have been studied extensively.\footnote{\, For a literature survey, see, {\em e.g.}, (Duarte {\em et al}, 2013). For a partial list (with some related literature, including on option pricing), see, {\em e.g.},
(Andersen {\em et al}, 2001),
(Anthony and MacDonald, 1998),
(Ayuso and Restoy, 1996),
(Ball and Roma, 1994),
(Bauer {\em et al}, 2009),
(Beetsma and Van Der Ploeg, 1994),
(Bekaert and Gray, 1998),
(Bertola and Caballero, 1992),
(Bertola and Svensson, 1993),
(Black and Scholes, 1973),
(Bo {\em et al}, 2011a, 2001b),
(Campa and Chang, 1996),
(Carr {\em et al}, 1998),
(Carr and Jarrow, 1990),
(Carr and Linetsky, 2000),
(Cavaliere, 1998),
(Chinn, 1991),
(Cornell, 2003),
(Christensen {\em et al}, 1998),
(De Jong, 1994),
(De Jong {\em et al}, 2001),
(Delgado and Dumas, 1992),
(Dominquez and Kenen, 1992),
(Driffill and Sola, 2006),
(Duarte {\em et al}, 2010),
(Dumas {\em et al}, 1995a, 1995b),
(Edin and Vredin, 1993),
(Edison {\em et al}, 1987),
(Flood and Garber, 1991),
(Flood {\em et al}, 1991),
(Garman and Kohlhagen, 1983),
(Grabbe, 1983),
(Harrison, 1985),
(Harrison and Pliska, 1981),
(Honogan, 1998),
(Hull and White, 1987),
(Kempa and Nelles, 1999),
(Klaster and Knot, 2002),
(Klein and Lewis, 1993),
(Koedijk {\em et al}, 1998),
(Krugman, 1991, 1992),
(Lai {\em et al}, 2008),
(Larsen and S{\o}rensen, 2007),
(Lin, 2008),
(Lindberg and S\"oderlind, 1994a, 1994b),
(Lindberg {\em et al}, 1993),
(Linetsky, 2005),
(Lundbergh and Ter\"asvirta, 2006),
(Magnier, 1992),
(McKinnon, 1982, 1984),
(Meese and Rose, 1990, 1991),
(Merton, 1973, 1976),
(Miller and Weller, 1991),
(Mizrach, 1995),
(Obstfeld and Rogoff, 1995),
(Rangvid and S{\o}rensen, 2001),
(Rose and Svensson, 1995),
(Saphores, 2005),
(Serrat, 2000),
(Slominski, 1994),
(Smith and Spencer, 1992),
(Sutherland, 1994),
(Svensson, 1991a, 1991b, 1992a, 1992b, 1993, 1994),
(Taylor and Iannizzotto, 2001),
(Torres, 2000a, 2000b),
(Tronzano {\em et al}, 2003),
(Veestraeten, 2008),
(Vlaar and Palm, 1993),
(Ward and Glynn, 2003),
(Werner, 1995),
(Williamson, 1985, 1986, 1987a, 1987b, 1989, 2002),
(Williamson and Miller, 1987),
(Yu, 2007),
(Zhang, 1994),
(Zhu, 1996),
and references therein.
}
Following Krugman (1991), the FX rate confined to a band with barriers is modeled as a stochastic process, where one needs to deal with the boundaries. There are essentially two choices: i) attainable boundaries, where the process is allowed to touch a boundary -- in this case the boundaries must be reflecting (see below); and ii) unattainable boundaries, where the process can get infinitesimally close to a boundary without ever touching it -- this is achieved by having the volatility of the process tend to zero (fast enough) as the process approaches a boundary. The unattainable boundary approach has been explored to a greater extent as dealing with reflecting boundaries can be tricky. However, with unattainable boundaries the underlying math typically is rather involved; {\em e.g.}, the pricing PDE for simple FX options (European call/put) either must be solved numerically or involves complicated special functions. Simply put, analytical tractability is challenging.

{}In this note we discuss -- in what is intended to be a pedagogical fashion -- FX option pricing in target zones with attainable boundaries. The basic idea behind option pricing in the presence of boundaries is no different than in the case without boundaries: we must construct a self-financing hedging strategy which replicates the claim at maturity. To do this, we must construct a discounted FX rate process and find a measure under which it is a martingale -- the risk neutral measure -- which is the requirement that there be no arbitrage. Then the option price is expressed via a conditional expectation of the discounted claim under this risk neutral measure, which leads to a Black-Scholes-like PDE. The key difference is that now, together with the terminal condition at maturity, we must also specify boundary conditions.

{}These boundary conditions must be reflecting, that is, they must be Neumann boundary conditions. This follows from the requirement that the identity process be a martingale under the risk neutral measure: simply put, the risk neutral measure must be normalized to 1 when summing over all possible outcomes, and this invariably forces reflecting boundary conditions. Put another way, if the boundary conditions are not reflecting, probability ``leaks" through the boundaries.

{}Reflecting boundary conditions imply that the differential short-rate -- the difference between the foreign and domestic short-rates -- cannot be constant; in fact, it cannot even be deterministic. This is a consequence of the requirements that: i) there be no arbitrage; ii) the FX rate be positive; and iii) the attainable boundaries be reflecting. Moreover, the requirement that the discounted FX rate be a martingale under the risk neutral measure fixes the differential short rate in terms of the functional form of the FX rate process as a function of the underlying Brownian motion together with the (generally, non-deterministic) drift and the volatility. This has a natural financial interpretation, to wit, as the Uncovered Interest Parity.

{}In most practical applications the width of the band is narrow\footnote{\, {\em E.g.}, the USD/HKD spread trades between 7.75 and 7.85, a band fixed by the Hong Kong Monetary Authority.}. This allows to take a pragmatic approach and pick the functional form of the FX rate process based on computational convenience. With a thoughtful choice, the FX option pricing problem can be solved analytically. In fact, the European call and put (and related) option prices are expressed via (fast converging) series of elementary (trigonometric) functions. We discuss the general approach to solving the pricing PDE and explicit examples. This includes analytically tractable models with (non-Ornstein-Uhlenbeck) mean-reversion, which are also solvable in elementary functions.

{}The remainder of this note is organized as follows. In Section \ref{sec3} we briefly review the general procedure for pricing FX options, with self-financing replicating strategies briefly reviewed in Appendix \ref{app.self-fin}. In Section \ref{sec4} we discuss pricing FX options in the presence of attainable reflecting boundaries, including hedging, European call and put options, explicit models, {\em etc.}, with some details relegated to Appendix \ref{app.0} and Appendix \ref{app.A}. We briefly conclude with some remarks in Section \ref{sec5}.

\section{FX Options}\label{sec3}

{}Let us assume that the domestic currency ({\em e.g.}, USD) is freely traded with no restrictions, whereas the foreign currency ({\em e.g.}, HKD) trades inside a target zone. We have a domestic cash bond $B^d_t$ and a foreign cash bond $B^f_t$. We also have the exchange rate $S_t$, which, for our purposes here, is the worth of one unit of the domestic currency in terms of the foreign currency ({\em e.g.}, in our USD/HKD example, $S_t$ is the HKD worth of 1 USD, whose target zone is 7.75 to 7.85). We will refer to tradables denominated in the foreign (domestic) currency as foreign (domestic) tradables. We are interested in pricing derivatives from a {\em foreign} investor's perspective. The foreign cash bond $B^f_t$ is a foreign tradable; however, $B^d_t$ and $S_t$ are not. We can construct another foreign tradable via
\begin{equation}
 {\widetilde S}_t \equiv B^d_t S_t
\end{equation}
(In our USD/HKD example above, this is the HKD value of the USD cash bond). The discounted process, which must be a martingale under the risk-neutral measure ${\bf Q}$ (see Appendix \ref{app.self-fin}), is given by
\begin{equation}\label{ZX}
 Z_t = (B^f_t)^{-1}{\widetilde S}_t = B_t^{-1} S_t
\end{equation}
where
\begin{equation}\label{eff.cash}
 B_t \equiv B^f_t / B^d_t
\end{equation}
The price of a claim $Y_T$ is given by (see Appendix \ref{app.self-fin})
\begin{equation}
 {\widetilde V}_t = B^f_t~\mathbb{E}\left((B^f_T)^{-1} Y_T\right)_{{\bf Q},{\cal F}_t}
\end{equation}
The foreign monetary authority, which confines the foreign currency to the target zone, (in theory) also adjusts the foreign interest rates based on the domestic interest rates and the FX rate. Therefore, we can assume that the domestic cash bond $B_t^f$ is deterministic within the (short enough) time horizons we are interested in for the purpose of pricing FX derivatives.\footnote{\, More generally, we can assume that any volatility in the domestic bond $B_t^d$ is uncorrelated with the volatility in the FX rate $S_t$ and the volatility in the foreign bond $B_t^f$, or, more precisely, any such correlation is negligible at relevant time horizons. This would not alter any of the subsequent discussions or conclusions, so for the sake of simplicity we will assume that $B_t^d$ is deterministic.}  For the claim price we then have
\begin{eqnarray}
 &&{\widetilde V}_t = B_t^d(B_T^d)^{-1}V_t\label{tildeV}\\
 &&V_t \equiv B_t~\mathbb{E}\left(B_T^{-1} Y_T\right)_{{\bf Q},{\cal F}_t}
\end{eqnarray}
Note that $B_t$ defined in (\ref{eff.cash}) is the ratio of the two cash bonds. We can define the corresponding differential (or ``effective") short-rate process via:
\begin{equation}
r_t \equiv {d\ln(B_t)\over dt} = r^f_t - r^d_t
\end{equation}
where $r^f_t$ and $r^d_t$ are the foreign and domestic short-rate processes:
\begin{eqnarray}
 &&r_t^f \equiv {d\ln(B^f_t)\over dt}\\
 &&r_t^d \equiv {d\ln(B^d_t)\over dt}
\end{eqnarray}
Note, however, that $r_t$ need not be positive. Also, here we are assuming that $r_t^d$ is deterministic; however, $r_t^f$ is not, nor is $r_t$. With this assumption, using (\ref{tildeV}), we can compute the actual price ${\widetilde V}_t$ of the claim $Y_T$ by computing the would-be ``price" $V_t$ of the claim $Y_T$ with $S_t$ and $B_t$ playing the roles of the tradable and the numeraire, respectively (see Appendix \ref{app.self-fin}). In the following, for the sake of notational and terminological convenience and brevity,\footnote{\, Alternatively, we can set $r_t^d$ to zero, so $B_t$ is the same as the foreign cash bond $B_t^f$, and restore the (deterministic) $r_t^d$ dependence at the end by multiplying all derivative prices by $B_t^d(B_T^d)^{-1}$.} we refer to $B_t$ as the cash bond, $r_t$ as the short-rate, and $V_t$ as the claim price; also, we refer to the FX rate $S_t$ as FXR.\footnote{\, FX has an analog in equities. Consider a stock with a continuous dividend rate $\delta_t$. Then the risk-free interest rate is analogous to the domestic short-rate, the dividend rate $\delta_t$ is analogous to the foreign short-rate, and the stock is analogous to the foreign currency (so the stock price is analogous to the worth of one unit of the foreign currency in terms of the domestic currency).}

\section{Pricing with Boundaries}\label{sec4}
\subsection{Martingales without Boundaries}

{}When we have no boundaries, typically we can construct a nontrivial martingale $Z_t$ other than the identity $I_t$. Thus, consider the transition density\footnote{\, {\em I.e.}, the probability density of starting from $W_t = z$ and ending at $W_{t^\prime} = z^\prime$, where $t^\prime > t$.}  for a ${\bf Q}$-Brownian motion $W_t$ (taking values on the entire real axis, $W_t\in {\bf R}$), which is the usual Gaussian distribution:
\begin{equation}
 P(t,z;t^\prime,z^\prime) = {1\over\sqrt{2\pi\left(t^\prime-t\right)}}~\exp\left(-{\left(z^\prime-z\right)^2\over 2\left(t^\prime-t\right)}\right)
\end{equation}
The identity $I_t$ is a martingale under this measure. However, there also exist other martingales, {\em e.g.}, $Z_t = W_t$ is a martingale, and so is
\begin{equation}\label{log.norm}
 Z_t \equiv S_0\exp\left(\sigma W_t - \sigma^2 t/2\right)
\end{equation}
where $S_0$ and $\sigma$ are constant.\footnote{\, In the log-normal Black-Scholes model the discounted process $B_t^{-1}S_t$ is given by (\ref{log.norm}).}

\subsection{Boundaries}

{}When boundaries are present, things are trickier. Thus, let us consider the process (here $W_t$ is a ${\bf Q}$-Brownian motion)
\begin{equation}\label{X}
 dX_t = \sigma(X_t) dW_t + \mu(X_t) dt
\end{equation}
where $\sigma(x)$ and $\mu(x)$ have no explicit time dependence.\footnote{\, We consider time-homogeneous dynamics so the problem is analytically tractable (see below).} In fact, for our purposes here, motivated by analytical tractability (see footnote \ref{foot.sigma}), it will suffice to consider constant $\sigma(x)\equiv\sigma$. However, for now we will keep $\mu(x)$ general (but Lipschitz continuous). We will now introduce barriers\footnote{\, With the view, {\em e.g.}, to have a function $S_t = f(X_t)$ with {\em attainable} barriers at $S_\pm = f(x_\pm)$. Also, note that, unless $\mu(x) \equiv 0$, this is not the same as having time-independent barriers for $W_t$.} for the process $X_t$ at $X_t = x_-$ and $X_t = x_+$ (see, e.g., \cite{Freidlin}). Below, without loss of generality, we will assume $x_- < x_+$.

{}We need to construct the measure ${\bf Q}$ under which the process $Z_t$ in (\ref{Z}) is a martingale. For our purposes here it will suffice to assume that\footnote{\, {\em I.e.}, i) $r_t$ is a {\em local} function of $X_t$ and $t$, and ii) $Y_T$ is independent of the history ${\cal F}_T$.} i) the short-rate $r_t = r(X_t, t)$ and ii) $Y_T = Y(X_T)$. Let us define the pricing function
\begin{equation}
 v(x, t, T) \equiv B_t~\mathbb{E}\left(B_T^{-1}Y_T\right)_{{\bf Q}, X_t = x}
\end{equation}
where $X_t$ is defined via (\ref{X}) (with constant $\sigma$). Note that $V_t$ in (\ref{E}) is given by $V_t = v(X_t, t, T)$. Since $E_t$ is a ${\bf Q}$-martingale, we have the following PDE for $v(x,t,T)$
\begin{equation}\label{PDE.v}
 \partial_t v(x,t,T) + \mu(x) \partial_x v(x,t,T) + {\sigma^2\over 2}\partial_x^2 v(x,t,T) - r(x,t) v(x,t,T) = 0
\end{equation}
subject to the terminal condition $v(x,T,T) = Y(x)$. Also, $r_t\equiv {d\ln(B_t)/dt}$.

{}Consider a ${\bf Q}$-martingale of the form $M_t = w(X_t, t)$, where $w(x,t)$ is a deterministic function. We have the following PDE:
\begin{equation}\label{PDE.w}
 \partial_t w(x,t,T) + \mu(x) \partial_x w(x,t,T) + {\sigma^2\over 2}\partial_x^2 w(x,t,T) = 0
\end{equation}
We must specify boundary conditions for $w(x,t)$ at $x=x_\pm$. For the identity $I_t$ to be a martingale under ${\bf Q}$, we must have reflecting (Neumann) boundary conditions\footnote{\, Thus, Dirichlet or Robin boundary conditions would be inconsistent with $I_t$ being a martingale. See Appendix \ref{app.0} for the transition density and martingales for Robin boundary conditions.}
\begin{equation}
 \partial_x w(x_\pm,t) = 0
\end{equation}
The same boundary conditions must be imposed on the pricing function $v(x,t,T)$:
\begin{equation}\label{b.v}
 \partial_x v(x_\pm,t,T) = 0
\end{equation}
Then the claim $Y(x)$ must satisfy the same boundary conditions:
\begin{equation}\label{b.Y}
 \partial_x Y(x_\pm) = 0
\end{equation}
which are consistent with the claim $Y(x)\equiv 1$ for a zero-coupon $T$-bond.

{}We can now show that $r_t$ cannot be deterministic. First, note that we wish our FXR process $S_t$ to stay within a band with {\em attainable} boundaries $S_\pm$. This can be achieved by having $S_t = f(X_t)$, where $f(x)$ is a bounded {\em monotonic}\footnote{\, Monotonicity is assumed so we can price claims (see below).} function on $[x_-,x_+]$ such that $f(x_\pm)=S_\pm$. In this regard, $S_t$ cannot have any explicit $t$ dependence,\footnote{\, Otherwise, barring any contrived time dependence, $S_t$ generically will break the band.} {\em i.e.}, $S_t$ depends on $t$ only via $X_t$. Second, since the discounted FXR process $Z_t = B_t^{-1}S_t$ is a ${\bf Q}$-martingale, the function $f(x)$ satisfies the same PDE as $v(x,t,T)$. It then follows that $r(x,t) = r(x)$, so the short-rate $r_t$ cannot have any explicit $t$ dependence either. We thus have the following ordinary differential equation for $f(x)$:
\begin{equation}\label{f}
 \mu(x) f^\prime(x) + {\sigma^2\over 2} f^{\prime\prime}(x) - r(x)f(x) = 0
\end{equation}
subject to the boundary conditions
\begin{equation}\label{b.f}
 f^\prime(x_\pm) = 0
\end{equation}
where $f^\prime(x) \equiv \partial_x f(x)$. We must also have $f(x) > 0$; therefore, $r(x)$ cannot be constant.\footnote{\, For constant $r_t>0$ ($r_t<0$) we would have minima (maxima) at $x=x_-$ and $x=x_+$ with a maximum (minimum) located between $x_-$ and $x_+$, which is not possible for $f(x)>0$ on $[x_-,x_+]$.} So, we can {\em choose} $f(x) > 0$ satisfying (\ref{b.f}) and view (\ref{f}) as fixing $r(x)$:
\begin{equation}\label{r}
 r(x) = \mu(x) {f^\prime(x)\over f(x)} + {\sigma^2\over 2} {f^{\prime\prime}(x)\over f(x)}
\end{equation}
This relation has a natural financial interpretation in the FX context (see below).

\subsection{Pricing PDE}

{}We can now tackle the pricing PDE (\ref{PDE.v}). Let
\begin{eqnarray}\label{g}
 && g(x) \equiv \ln(f(x))\\
 && u(x,t,T)\equiv \exp\left({1\over\sigma^2}\int_{x_-}^x dy~\mu(y)\right) v(x,t,T)\label{u.1}
\end{eqnarray}
Then, taking into account (\ref{r}), we have:
\begin{equation}
 \partial_t u(x,t,T) + {\sigma^2\over 2}\left[\partial_x^2 u(x,t,T) - U(x) u(x,t,T)\right]= 0
\end{equation}
where the ``potential" $U(x)$ is given by
\begin{eqnarray}\label{U}
 &&U(x) \equiv h^2(x) + h^\prime(x)\\
 &&h(x)\equiv g^\prime(x)+ {\mu(x)\over\sigma^2}\label{h}
\end{eqnarray}
subject to the boundary and terminal conditions (note that $g^\prime(x_\pm)=0$ due to (\ref{b.f}))
\begin{eqnarray}\label{b.u}
 &&\partial_x u(x_\pm, t, T) = h(x_\pm) u(x_\pm, t, T)\\
 &&u(x,T,T) = \exp\left({1\over\sigma^2}\int_{x_-}^x dy~\mu(y)\right) Y(x)\label{term}
\end{eqnarray}
We have standard separation of variables and the solution is given by\footnote{\, We assume that $U(x)$ is bounded on $[x_-,x_+]$, so the spectrum $E_n$ is bounded from below.}
\begin{equation}\label{u.c}
 u(x,t,T) = \sum_{n=0}^\infty c_n~\psi_n(x)~e^{-E_n\left(T-t\right)}
\end{equation}
where $\psi_n(x)$ form a complete orthonormal set of solutions to the static Schr\"odinger equation ($\delta_{n n^\prime}$ is the Kronecker delta)
\begin{eqnarray}\label{psi}
 &&-{\sigma^2\over 2} \left[\psi^{\prime\prime}_n(x) - U(x)\psi_n(x)\right] = E_n\psi_n(x)\\
 &&\int_{x_-}^{x_+} dx~\psi_n(x)~\psi_{n^\prime}(x) = \delta_{n n^\prime}\label{ortho}
\end{eqnarray}
subject to the boundary conditions\footnote{\, Notice that $-{2\over\sigma^2}\left(E_n - E_{n^\prime}\right)\int_{x_-}^{x_+} dx~\psi_n(x)~\psi_{n^\prime}(x) = \int_{x_-}^{x_+} dx\left[\partial_x^2\psi_n(x)~\psi_{n^\prime}(x) - (n\leftrightarrow n^\prime) \right] = \left.\left[\partial_x\psi_n(x)~\psi_{n^\prime}(x) - (n\leftrightarrow n^\prime) \right]\right|_{x_-}^{x_+} = 0$ by virtue of (\ref{b.psi}), hence (\ref{ortho}) for $n\neq n^\prime$ as $E_n\neq E_{n^\prime}$.}
\begin{equation}\label{b.psi}
 \psi^\prime_n(x_\pm) = h(x_\pm)\psi_n(x_\pm)
\end{equation}
As above, $\psi^\prime_n(x) \equiv \partial_x\psi_n(x)$.

{}The coefficients $c_n$ in (\ref{u.c}) are fixed using the terminal condition (\ref{term}) and (\ref{ortho}):
\begin{equation}
 c_n = \int_{x_-}^{x_+} dx^\prime \exp\left({1\over\sigma^2}\int_{x_-}^{x^\prime} dy~\mu(y)\right)\psi_n\left(x^\prime\right)Y(x^\prime)
\end{equation}
The spectrum $E_n$ is nonnegative. For the eigenfunction ($a_0$ is fixed via (\ref{ortho}))
\begin{eqnarray}\label{psi.0}
 &&\psi_0(x) \equiv a_0\exp\left(\int_{x_-}^x dy~h(y)\right)\\
 &&a_0 \equiv \left[\int_{x_-}^{x_+} dx^\prime~e^{2\int_{x_-}^{x^\prime}dy~h(y)}\right]^{-{1\over 2}}
\end{eqnarray}
we have $E_0 = 0$. The other eigenvalues $E_1 < E_2 <\dots$ are all positive.\footnote{\, Indeed, from (\ref{ortho}) with $n>0$ and $n^\prime=0$ and the fact that $\psi_0(x) > 0$, it follows that $\psi_n(x)$ must flip sign on $[x_-,x_+]$, {\em i.e.}, $\psi_n(x)$ has at least one node. However, if any $E_n < 0$, then $\psi_{n_*>0}(x)$ corresponding to $E_{n_*} \equiv \mbox{min}(E_n) < 0$ would have to have at least one node, which is not possible.}

{}Putting everything together, we get the following formula for the pricing function:
\begin{eqnarray}\label{deriv.prc}
 &&v(x,t,T) = f(x) \left[{\widetilde c}_0 + {1\over\psi_0(x)} \sum_{n=1}^\infty {\widetilde c}_n~\psi_n(x)~e^{-E_n\left(T-t\right)}\right]\\
 &&{\widetilde c}_n \equiv \int_{x_-}^{x_+} dx^\prime~\psi_0\left(x^\prime\right)\psi_n\left(x^\prime\right) {Y(x^\prime)\over f(x^\prime)},~~n\geq 0
\end{eqnarray}
When $Y(x) = f(x)$, {\em i.e.}, $Y_T = S_T$, we have ${\widetilde c}_0=1$ and ${\widetilde c}_{n>0} = 0$, so $v(x,t,T) = f(x)$, as it should be since this is simply the pricing function for a forward.

\subsection{Hedging}

{}Above we imposed the boundary conditions (\ref{b.f}) on the FXR process. This implies that the local FXR volatility vanishes at the boundaries. However, unlike the case of unattainable boundaries, here the boundaries are {\em attainable}: the FXR process touches a boundary and is reflected back into the band. Furthermore, note that in any finite period, $S_t$ can touch a boundary multiple (unbounded number of) times.

{}Even though the local FXR volatility vanishes at the boundaries, the hedging strategy is well-defined. The number of the $S_t$ units held by the hedging strategy\footnote{\, The actual hedge (recall that we are operating from the foreign investor's perspective) consists of holding $\phi_t$ units of the domestic cash bond, and $\psi_t$ units of the foreign cash bond (which is the foreign investor's numeraire). As mentioned above, if we set the domestic short-rate to zero, $S_t$ (the worth of 1 unit of the domestic cash bond in the foreign currency) becomes a foreign tradable.}
\begin{equation}\label{phi.hedge}
 \phi_t = {\partial V_t\over \partial S_t} = {\partial_x v(x, t, T)\over f^\prime(x)}
\end{equation}
Since we have (\ref{b.v}), so long as $f^{\prime\prime}(x_\pm)$ are finite, $\phi_t$ is finite at the boundaries:
\begin{equation}
 \left.\phi_t\right|_{x = x_\pm} = {\partial^2_x v(x_\pm, t, T)\over f^{\prime\prime}(x_\pm)}
\end{equation}
Consequently, the cash bond holding $\psi_t$ is also well-defined at the boundaries.

\subsection{Call and Put}

{}Consider claims of the form $Y^c_T(k) = (S_T - k)^+ = \mbox{max}(S_T - k, 0)$ (European call option with maturity $T$ and strike $k$) and $Y^p_T(k) = (k - S_T)^+ = \mbox{max}(k - S_T, 0)$ (European put option with maturity $T$ and strike $k$). We have
\begin{equation}
 Y^c_T(k) - Y^p_T(k) = Y^f_T(k)
\end{equation}
where $Y^f_T(k) = S_T - k$ is the claim for a forward with maturity $T$ and ``strike" $k$. We have the usual put-call parity: $V^c_t(k,T)-V^p_t(k,T) = V^f_t(k,T)$. The forward price
\begin{equation}
 V^f_t(k,T) = S_t - k~P(t,T)
\end{equation}
where $P(t,T) \equiv v^{bond}({\widehat x}_t,t,T)$ is a zero-coupon $T$-bond price (with $Y^{bond}_T = 1$), and the call price
\begin{eqnarray}\label{gen-call}
 &&V^c_t(k,T) = S_t \left[{\widetilde c}_0 + {1\over\psi_0({\widehat x}_t)} \sum_{n=1}^\infty {\widetilde c}_n~\psi_n({\widehat x}_t)~e^{-E_n\left(T-t\right)}\right]\\
 &&{\widetilde c}_n \equiv \int_{x_*}^{x_+} dx^\prime~\psi_0\left(x^\prime\right)\psi_n\left(x^\prime\right) \left[1 - {k\over f(x^\prime)}\right],~~n\geq 0
\end{eqnarray}
where\footnote{\, This is where the monotonicity of $f(x)$ is important.} $f(x_*)\equiv k$, $f({\widehat x}_t) \equiv S_t$ ($S_- < S_+$, $S_\pm \equiv f(x_\pm)$). For the binary option $Y^b_T = \theta(S_T-k)$ and $V_t^b(k, T) = -\partial V^c_t(k,T)/\partial k$ ($\theta(y)$ is the Heaviside step-function).

\subsection{FX Rate Process}

{}In most practical applications the band is narrow, so we can choose the FXR process based on computational convenience. Note that $\psi_1(x)$ has one node $x_1$ on $[x_-,x_+]$: $\psi_1(x_1) = 0$. In the cases where $\psi_1(x)/\psi_0(x)$ is a monotonic function on $[x_-,x_+]$, we can choose the FXR process as follows (note that $f^\prime(x_\pm) = 0$):
\begin{equation}\label{FXR.pr}
 f(x) = S_{mid}\left[1 + \gamma~{\psi_1(x)\over\psi_0(x)}\right]^{-1}
\end{equation}
Here $S_{mid} \equiv f(x_1)$. Without loss of generality we can assume $\psi_1(x_-)>0$ and $\psi_1(x_+) < 0$, so we have $S_- < S_+$ for $\gamma > 0$. Since the band is narrow, $\gamma\ll 1$.

{}For the FXR process (\ref{FXR.pr}) the zero-coupon $T$-bond price $P(t,T)$, for which $Y(x)\equiv 1$, simplifies. For this claim ${\widetilde c}_0=1/S_{mid}$, ${\widetilde c}_1 = \gamma/S_{mid}$ and ${\widetilde c}_{n>0} = 0$, so we have
\begin{equation}\label{T-bond}
 P(t,T) = {S_t\over S_{mid}} + \left[1 - {S_t\over S_{mid}}\right] e^{-E_1\left(T-t\right)}
\end{equation}
Note that $P(t,T)$ can be greater than 1 as the short-rate $r(x)$ need not be positive (recall that $r(x)$ is the differential short-rate). More on this below.

\subsection{Explicit Models}

{}We have two functions: $g(x)$ and $\mu(x)$. If we set $\mu(x) = 0$ (or some other constant), $X_t$ is a Brownian motion (with a constant drift). Then $U(x)$ is not that simple, albeit still tractable. Alternatively, we can take $g(x)$ and $\mu(x)$ such that $U(x)=0$.

\subsubsection{Vanishing Drift}

{}Let $\mu(x)\equiv 0$. Also, let $x_- = 0$, $x_+ = L$. Then we can take (note that this choice differs from (\ref{FXR.pr}))
\begin{equation}\label{quart1}
 g(x) = \gamma\left[3~{x^2\over L^2} - 2~{x^3\over L^3}\right] + \ln(S_-)
\end{equation}
where $\gamma>0$, so that $S_+ = S_-\exp(\gamma)$. We have a quartic potential
\begin{equation}\label{quartic}
 U(x) = {6\gamma\over L^2}\left[1-2~{x\over L} + 6\gamma\left({x\over L} - {x^2\over L^2}\right)^2\right]
\end{equation}
which is well-studied using perturbation theory. However, since $\gamma\ll 1$, we have simplifications. Recall that $E_0=0$ irrespective of $\gamma$. Also, $\psi_0(x) = a_0 f(x)/ f(0)\approx a_0 \approx 1/\sqrt{L}$. In the zeroth approximation, {\em i.e.}, in the limit $\gamma\rightarrow 0$ where $U(x)\rightarrow 0$, we have $E^{(0)}_n = \pi^2 n^2 \sigma^2/2L^2$ (see below). For the $n>0$ levels the corrections due to nonzero $\gamma$ are controlled by the ratio
\begin{equation}
 {\sigma^2 U(x)\over 2{E^{(0)}_n}} = {6\gamma\over \pi^2 n^2}\left[1-2~{x\over L} + 6\gamma\left({x\over L} - {x^2\over L^2}\right)^2\right]
\end{equation}
which is small for $\gamma\ll 1$. We can therefore set $U(x)\approx 0$. If we wish to account for the leading ${\cal O}(\gamma)$ corrections, we can drop the nonlinear term in (\ref{quartic}), which gives a linear potential
\begin{equation}
 U(x) \approx {6\gamma\over L^2}\left[1-2~{x\over L}\right]
\end{equation}
for which the solutions to the Schr\"odinger equation (\ref{psi}) are expressed in terms of the Airy functions $Ai(x)$ and $Bi(x)$. Alternatively, we can use the WKB approximation.

\subsubsection{Vanishing Potential}

{}We can set the potential $U(x)$ to zero without any approximations at the ``expense" (see below) of having nonvanishing $\mu(x)$:
\begin{equation}\label{mu.g}
 \mu(x) = -\sigma^2 g^\prime(x)
\end{equation}
Then, setting $x_-=0$ and $x_+=L$, irrespective of $g(x)$, we have $\psi_0(x) = 1/\sqrt{L}$ ($E_0 = 0$), and for $n>0$
\begin{eqnarray}
 &&\psi_n(x) = \sqrt{2\over L} \cos\left({\pi n x\over L}\right)\\
 &&E_n = {\pi^2 n^2\sigma^2\over 2L^2}\label{E_n}
\end{eqnarray}
The call price simplifies to
\begin{eqnarray}\label{call.cos}
 &&V^c_t(k,T) = S_t \left[{\widetilde c}_0 + \sqrt{L} \sum_{n=1}^\infty {\widetilde c}_n~\psi_n({\widehat x}_t)~e^{-E_n\left(T-t\right)}\right]\\
 &&{\widetilde c}_n = {1\over\sqrt{L}} \int_{x_*}^L dx^\prime~\psi_n\left(x^\prime\right) \left[1 - {k\over f(x^\prime)}\right],~~n\geq 0
\end{eqnarray}
where $f(x_*)\equiv k$, $f({\widehat x}_t) \equiv S_t$. Since $\psi_1(x)$ is monotonic on $[0,L]$, it is convenient to take $f(x)$ of the form (\ref{FXR.pr}):
\begin{equation}\label{cos}
 f(x) = S_{mid}\left[1 + \sqrt{2}\gamma\cos\left(\pi x\over L\right)\right]^{-1}
\end{equation}
We then have ($S_-\leq k\leq S_+$, $S_\pm = S_{mid}/(1\mp\sqrt{2}\gamma)$)
\begin{eqnarray}
 &&{\widetilde c}_0 = {\sqrt{2}\gamma k\over \pi S_{mid}}\left[(\pi - \phi_*)\cos(\phi_*) + \sin(\phi_*)\right]\label{c.0}\\
 &&{\widetilde c}_1 = -{\gamma k\over \pi S_{mid}}\left[\pi - \phi_* + \sin(\phi_*)\cos(\phi_*)\right]\label{c.1}\\
 &&{\widetilde c}_{n>1} = {\gamma k\over \pi S_{mid}}\left[{\sin((n+1)\phi_*)\over{n+1}} + {\sin((n-1)\phi_*)\over{n-1}}  - {2\cos(\phi_*)\sin(n\phi_*)\over n}\right]\label{c.n}\\
 &&\phi_*\equiv \arccos\left({1\over\sqrt{2}\gamma}\left[{S_{mid}\over k} - 1\right]\right)
\end{eqnarray}
In practical computations we would truncate the series in (\ref{call.cos}) at suitable finite $n$. Also, note that the $T$-bond price is given by (\ref{T-bond}).

{}Here the following remark is in order. As mentioned above, the local FXR volatility vanishes at the boundaries, which is due to (\ref{b.f}). There is no way around this: boundaries must be reflecting, and then we must have (\ref{b.f}). A simple way to see this is that otherwise (\ref{b.Y}) will not be satisfied for claims such as call $Y^c(x) = (f(x) - k)^+$ and put $Y^p(x) = (k - f(x))^+$, {\em i.e.}, we would not be able to hedge such claims. That the local FXR volatility vanishes at the boundaries in itself is not problematic. In fact, $p(x)\equiv g^\prime(x) = f^\prime(x)/f(x)$ is small compared with its maximal value $p_{max}$ on $[x_-,x_+]$ only in relatively small regions adjacent to the boundaries. Thus, in the model (\ref{cos}) we have
\begin{equation}\label{p}
 p(x) = {\sqrt{2}\gamma\over {1+\sqrt{2}\gamma\cos(\pi x/L)}} {\pi\sin(\pi x/L)\over L}\approx p_{max}\sin(\pi x/L)
\end{equation}
where $p_{max}\approx p(L/2) = \sqrt{2}\gamma\pi/L$, and we have taken into account that $\gamma\ll 1$. So, $p(L/6) \approx 0.5~p_{max}$, $p(L/10)\approx 0.31~ p_{max}$, {\em etc.}, {\em i.e.}, due to the nonlinearity of $p(x)$, even at $x=L/10$ the local FXR volatility is not too small (compared with $p_{max}$).\footnote{\, In the model (\ref{quart1}) we have $p(x)=4p_{max}(x/L - (x/L)^2)$, where $p_{max}=p(L/2)=3\gamma/2L$, so $p(x/10)=0.36~p_{max}$.}

\subsubsection{Nonvanishing Potential and Drift}

{}One ``shortcoming" of the model (\ref{cos}) is that, since we have (\ref{mu.g}), the drift $\mu(x) = -\sigma^2 p(x)$ is negative away from the boundaries, albeit it is small (compared with $\sqrt{2}\sigma^2\pi/L$) as it is suppressed by $\gamma\ll 1$. Therefore, in a long run, on average $X_t$ will slowly drift toward 0. This can be circumvented by considering models where both the potential $U(x)$ and the drift $\mu(x)$ are nonvanishing. Since $\gamma\ll 1$, with the appropriate choice of $h(x)$, to the leading order $\mu(x)\approx \sigma^2 h(x)$. Alternatively, we can take the desired drift and treat the terms in the potential stemming from $g^\prime(x)$ in (\ref{h}) as small. For our purposes here, the former approach is more convenient.

{}Thus, one evident choice is $h(x) = \alpha(\theta - x)/\sigma^2$, where $\theta$ and $\alpha$ are constant. Then $\mu(x)\approx \alpha(\theta - x)$, so $X_t$ (approximately) follows the mean-reverting Ornstein-Uhlenbeck (OU) (Uhlenbeck and Ornstein, 1930) process, $\alpha$ is the mean-reversion parameter, and $\theta$ is the long-run expected value of $X_t$ (which we can set to $L/2$). In this case we have a quadratic potential and $\psi_n(x)$ in (\ref{psi}) are expressed via the parabolic cylinder functions.

{}However, with an appropriate choice of $h(x)$, we can also have a solution expressed purely via elementary functions. Thus, consider
\begin{equation}\label{tan}
 h(x) = -\nu\tan\left(\nu\left[x - {L\over 2}\right]\right)
\end{equation}
where $0 < \nu < \pi/L$ is a constant parameter. Then we have a constant potential $U(x) \equiv -\nu^2$. The eigenfunctions $\psi_n(x)$ read ($n=0,1,2,\dots$):
\begin{eqnarray}
 &&\psi_n(x) = a_n\cos\left(\lambda_n\left[x - {L\over 2}\right] + {\pi n\over 2}\right)\\
 &&a_n = \left({L\over 2}\left[1 + (-1)^n{\sin(\lambda_n L)\over\lambda_n L}\right]\right)^{-1/2}\label{a.n.tan}
\end{eqnarray}
where $\lambda_n$ are the positive roots of the following equation (which follows from (\ref{b.psi})):
\begin{equation}\label{lambda.n.tan}
 \lambda_n\tan\left([\lambda_n L - \pi n]/2\right) = \nu\tan(\nu L/2)
\end{equation}
The smallest root is $\lambda_0=\nu$, and $\lambda_0<\lambda_1<\lambda_2<\dots$ (Note that $E_n = \sigma^2(\lambda_n^2-\nu^2)/2$.)

{}The call option price is given by (\ref{gen-call}) with ${\widetilde c}_n$ defined in Appendix \ref{app.A}. The zero-coupon $T$-bond price is given by (\ref{T-bond}). If $\nu\sim\pi/L$ (although recall that $\nu<\pi/L$),\footnote{\, In fact, here we assume that $\nu$ is not too close to $\pi/L$ or else the drift becomes large near the boundaries. In the $\nu\rightarrow \pi/L$ limit the boundaries are no longer attainable.} then, assuming $\gamma \ll 1$, the drift $\mu(x)\approx-\sigma^2\nu\tan(\nu(x-L/2))$ and we have positive drift for $x<L/2$ and negative drift for $x>L/2$, so we have a mean-reverting behavior.\footnote{\, Near $x=L/2$ the drift is approximately linear as in the OU process: $\mu(x)\approx \sigma^2\nu^2(L/2-x)$; however, away from the long-run value ({\em i.e.}, $L/2$), the nonlinear effects become important. Unlike the OU case with reflecting boundaries, the model (\ref{tan}) is solvable via elementary functions.} The $g^\prime(x)$ contribution into $\mu(x)$ via (\ref{h}) introduces a small asymmetry into $\mu(x)$ but does not alter the qualitative picture.

\subsection{Differential Rate}

{}The meaning of (\ref{r}), which stems from the requirement that there be no arbitrage ({\em i.e.}, that the discounted process (\ref{ZX}) be a martingale under the risk-neutral measure ${\bf Q}$), has a natural financial interpretation as the Uncovered Interest Parity. To illustrate this, let us momentarily step away from the target zone case and consider the case where neither the domestic nor the foreign currencies are constrained in any way. Then, if we take a familiar ``log-normal" form for the FX rate via $S_t = \exp(X_t)$, from (\ref{r}) we have
\begin{equation}\label{UIP}
 r_t = r^f_t - r^d_t = \mu(X_t) + {\sigma^2\over 2}
\end{equation}
Recalling that $dX_t = \sigma dW_t + \mu(X_t)dt$, (\ref{UIP}) is indeed the Uncovered Interest Parity.\footnote{\, The $\sigma^2/2$ shift is due to the log-normal form of the FX rate. {\em E.g.}, if $\mu(X_t) \equiv \mu=\mbox{const.}$, the expectation $\mathbb{E}\left(S_T\right)_{{\bf Q},{\cal F}_t} = \exp((\mu+\sigma^2/2)(T-t))$.}

{}In fact, there is a simple formula for the differential short-rate $r_t$. Using (\ref{r}), (\ref{g}), (\ref{h}), (\ref{psi}), (\ref{psi.0}) and (\ref{FXR.pr}), we have
\begin{equation}\label{r.1}
 r_t = E_1\left[1 - {f(X_t)\over S_{mid}}\right] = E_1\left[1 - {S_t\over S_{mid}}\right]
\end{equation}
so $r_t$ is positive (negative) at the lower (upper) barrier, which is a consequence of the requirement that there be no arbitrage.\footnote{\, If the domestic short-rate is low, $r_t^d < E_1\left(S_+/S_{mid} - 1\right)$, then the foreign short-rate would become negative for $S_t > S_{mid}\left(1 + r_t^d / E_1\right)$. Theoretically this would appear to imply arbitrage. However, in practice the short-rate is not a tradable instrument and this situation may not be arbitrageable as the tradable bonds (for the actually available maturities) may still have well-behaved yields despite the negative underlying short-rate (for a recent discussion, see, {\em e.g.}, (Kakushadze, 2015) and references therein). Also note that (\ref{r.1}) simply follows from $r_t = f(t,t)$ and (\ref{T-bond}), where $f(t, T) = -\partial_T \ln\left(P(t,T)\right)$ is the forward rate.}  Note that (\ref{r.1}) does not explicitly depend on $U(x)$ so long as we have (\ref{FXR.pr}).\footnote{\, And $r_t$ is continuous even if $U(x)$ contains $\delta$-functions, {\em i.e.}, when $\mu(x)$ is discontinuous.}

\section{Concluding Remarks}\label{sec5}

{}As we saw above, with a thoughtful choice of the FX rate process -- which choice, from a practical viewpoint, exists because the band is narrow and said choice does not affect quantitative results much -- we can solve the FX option pricing problem in the target zone analytically, in fact, via elementary functions. This is assuming attainable barriers.\footnote{\, Hence our choice of a constant diffusion coefficient $\sigma(x)\equiv\sigma$. When the band is narrow, there is little benefit to having nonconstant $\sigma(x)$ as the boundaries are reflecting. This is to be contrasted with the case of unattainable boundaries where $\sigma(x)$ tends to zero near the boundaries.\label{foot.sigma}} For unattainable barriers the math typically is more involved. Also, in practice the exchange rates in target zones frequently attain the boundaries, so attainable boundaries are also appealing from this viewpoint. In fact, in some cases the FX options markets imply a future expectation that the FX rate will break the band. Models accommodating band breaking are outside of the scope of this note; however, the fact that the markets sometimes price band breaks further indicates appealability of models with attainable boundaries.

\subsection*{Acknowledgments}
{}ZK would like to thank Eyal Neuman for discussions on Brownian motion with reflecting barriers, which prompted him to think about this topic. We would also like to thank Travis Fisher for discussions.

\appendix

\section{Self-financing Hedging Strategies}\label{app.self-fin}

{}The following discussion is rather general and applies to a wide class of underlying tradable instruments. So, suppose we have a tradable $S_t$ and a numeraire\footnote{\, Usually, the numeraire is chosen to be a cash bond, but it can be any tradable instrument.} $B_t$. Generally, $B_t$ need not be deterministic. Consider a claim $Y_T$ at the maturity time $T$. We wish to value this claim at times $t < T$. To do this, we need to construct a self-financing hedging strategy which replicates the claim $Y_T$. The hedging strategy amounts to, at any given time $t$, holding a portfolio $(\phi_t,\psi_t)$ consisting of $\phi_t$ units of $S_t$ and $\psi_t$ units of $B_t$, where $\phi_t$ and $\psi_t$ are previsible processes. The value $V_t$ of this portfolio at time $t$ is given by
\begin{equation}\label{V}
 V_t = \phi_t S_t + \psi_t B_t
\end{equation}
The self-financing property means that the change in the value of the portfolio is solely due the changes in the values of $S_t$ and $B_t$, {\em i.e.}, there is no cash flowing in or out of the strategy at any time:
\begin{equation}\label{self}
 dV_t = \phi_t dS_t + \psi_t dB_t
\end{equation}
Then from (\ref{self}) it follows that
\begin{equation}\label{E.Z}
 dE_t = \phi_t dZ_t
\end{equation}
where
\begin{eqnarray}
 &&E_t \equiv B_t^{-1} V_t\label{E}\\
 &&Z_t \equiv B_t^{-1}S_t\label{Z}
\end{eqnarray}
So, (\ref{E.Z}) relates the discounted claim price $E_t$ to the discounted tradable price $Z_t$.

{}Let us now assume that we can construct a measure ${\bf Q}$ under which $Z_t$ is a martingale. Then we can construct a self-financing strategy which replicates the claim $Y_T$ by setting (${\cal F}_t$ is the filtration up to time $t$, and $\mathbb{E}(\cdot)$ denotes expectation)
\begin{equation}
 E_t = \mathbb{E}\left(B_T^{-1} Y_T\right)_{{\bf Q},{\cal F}_t}
\end{equation}
We then have $V_T = B_T E_T = Y_T$. Since both $E_t$ and $Z_t$ are ${\bf Q}$-martingales, pursuant to the martingale representation theorem $\phi_t$ is a previsible process. Furthermore, from (\ref{V}), (\ref{E}) and (\ref{Z}) we have
\begin{equation}\label{psi.bond}
 \psi_t = E_t - \phi_t Z_t
\end{equation}
so $\psi_t$ is also previsible.

{}In the applications of the above discussion in the main text, we assume\footnote{\, Otherwise, the market would be incomplete and we would not be able to hedge claims.} that a single ${\bf Q}$-Brownian motion $W_t$ underlies the dynamics of $S_t$ and $B_t$. We also assume that the identity $I_t\equiv 1$ is a ${\bf Q}$-martingale, {\em i.e.}, $\mathbb{E}\left(I_T\right)_{{\bf Q}, {\cal F}_t} = I_t = 1$, so the measure ${\bf Q}$ is properly normalized when summed over all final outcomes at time $T$ irrespective of the history ${\cal F}_t$ prior to time $t < T$.

\section{Transition Density}\label{app.0}

{}Here we give the transition density for the process ($W_t$ is a ${\bf P}$-Brownian motion)
\begin{equation}
 dX_t = \sigma~dW_t + \mu~dt
\end{equation}
where $\sigma$ and $\mu$ are constant, and $X_t$ is allowed to wander between two boundaries at $X_t = x_-$ and $X_t = x_+$. Without loss of generality we can set $x_-=0$ and $x_+=L$.

{}Let $Y_T \equiv Y(X_T)$ be a claim, where $Y(x)$ is a continuous function, and let
\begin{equation}
  w(x, t, T) \equiv \mathbb{E}\left( Y(X_T)\right)_{{\bf P},X_t=x}
\end{equation}
Since $Z_t\equiv w(X_t, t, T)$ is a ${\bf P}$-martingale, $w(x,t,T)$ satisfies the following PDE
\begin{equation}\label{w1}
 \partial_t w(x,t,T) + \mu\partial_x w(x,t,T) + {\sigma^2\over 2}\partial_x^2 w(x,t,T) = 0
\end{equation}
subject to the terminal condition
\begin{equation}\label{term.app}
 w(x, T, T) = Y(x)
\end{equation}
Also, we must specify the boundary conditions at $x=0$ and $x=L$. We will impose the Robin boundary conditions:\footnote{\, Here, more generally, we can impose different Robin boundary conditions at $x=0$ and $x=L$. For our purposes here it will suffice to consider (\ref{bpm}). Let us mention that, in the case of different boundary conditions the spectrum generally has an infinite tower of positive eigenvalues, and also two additional eigenvalues, at least one of which is negative. (More precisely, there are non-generic degenerate cases with only one such additional eigenvalue, which is negative.)}
\begin{equation}
 \partial_x w(x_\pm, t, T) = \rho w(x_\pm, t, T)\label{bpm}
\end{equation}
which imply that the claim also satisfies the same boundary conditions:
\begin{equation}
 \partial_x Y(x_\pm) = \rho Y(x_\pm)\label{Ypm}
\end{equation}
Here $\rho$ is constant. When $\rho=0$ we have Neumann boundary conditions (so the boundaries are reflecting), while when $\rho\rightarrow\infty$ we have Dirichlet boundary conditions (so the boundaries are absorbing).

{}Let the probability density of starting at $X_t = x$ at time $t$ and ending at $X_{t^\prime} = x^\prime$ at time $t^\prime$ be $P(t,x;t^\prime,x^\prime)$, a.k.a. transition density. Since the claim $Y(X_T)$ depends only on the final value $X_T$, we have
\begin{equation}
 w(x,t,T) = \int_0^L dx^\prime~P(t,x;T;x^\prime)~Y(x^\prime)
\end{equation}
So, $P(t,x;t^\prime,x^\prime)$ is a Green's function (a.k.a. heat kernel). The transition density can be computed using the eigenfunction method (see, e.g., \cite{Linetsky1}) and is given by:
\begin{eqnarray}
 &&P(t,x;T,x^\prime) = 2{\widetilde\rho}~{e^{\rho x + \left(2{\widetilde\rho}-\rho\right)x^\prime}\over{e^{2L{\widetilde\rho}} - 1}}~ e^{-E_0\left(T-t\right)}+
 \nonumber\\
 &&+{2\over L}~e^{\left({\widetilde\rho}-\rho\right)\left(x^\prime-x\right)} \sum_{n=1}^\infty {e^{-E_n\left(T-t\right)}\over q_n} \left[\pi n \cos\left({\pi n x\over L}\right) + L{\widetilde\rho}\sin\left({\pi n x\over L}\right)\right]\times\nonumber\\
 &&\times
 \left[\pi n \cos\left({\pi n x^\prime\over L}\right) + L{\widetilde\rho}\sin\left({\pi n x^\prime\over L}\right)\right]\label{tr.dens}
\end{eqnarray}
where
\begin{eqnarray}
 &&E_0 = \rho\left(\rho-2{\widetilde\rho}\right){\sigma^2\over 2}\\
 &&E_n = E_0 + {q_n\sigma^2\over 2L^2}\\
 &&q_n \equiv \pi^2 n^2 + L^2{\widetilde\rho}^2\\
 &&{\widetilde\rho} \equiv \rho + {\mu\over \sigma^2}
\end{eqnarray}
Let us assume $\rho\neq 0$. Then $E_0 = 0$ if ${\widetilde\rho} = \rho/2$, {\em i.e.},
\begin{equation}
 \rho = -{2\mu\over\sigma^2}
\end{equation}
So, we have
\begin{eqnarray}
 &&P(t,x;T,x^\prime) = \rho~{\exp\left(\rho x\right)\over{\exp\left(L\rho\right) - 1}} +
 \nonumber\\
 &&+{2\over L}~\exp\left({\rho\over 2}\left[x-x^\prime\right]\right)\sum_{n=1}^\infty {e^{-E_n\left(T-t\right)}\over q_n} \left[\pi n \cos\left({\pi n x\over L}\right) + {L\rho\over 2}\sin\left({\pi n x\over L}\right)\right]\times\nonumber\\
 &&\times
 \left[\pi n \cos\left({\pi n x^\prime\over L}\right) + {L\rho\over 2}\sin\left({\pi n x^\prime\over L}\right)\right]\label{tr.dens.1}
\end{eqnarray}
where
\begin{eqnarray}
 &&E_n = {q_n\sigma^2\over 2L^2}\\
 &&q_n = \pi^2 n^2 + {L^2\rho^2\over 4}
\end{eqnarray}
Under the measure (\ref{tr.dens.1}), the process
\begin{equation}
 Z_t = S_0~\exp(\rho X_t)
\end{equation}
is a martingale; however, the identity process $I_t$ is not.

{}On the other hand, when $\rho = 0$, we have
\begin{eqnarray}
 &&P(t,x;T,x^\prime) = 2{\widetilde\rho}~{e^{2{\widetilde\rho}x^\prime}\over{e^{2L{\widetilde\rho}} - 1}}+
 \nonumber\\
 &&+{2\over L}~e^{{\widetilde\rho}\left(x^\prime-x\right)} \sum_{n=1}^\infty {e^{-E_n\left(T-t\right)}\over q_n} \left[\pi n \cos\left({\pi n x\over L}\right) + L{\widetilde\rho}\sin\left({\pi n x\over L}\right)\right]\times\nonumber\\
 &&\times
 \left[\pi n \cos\left({\pi n x^\prime\over L}\right) + L{\widetilde\rho}\sin\left({\pi n x^\prime\over L}\right)\right]\label{tr.dens.2}
\end{eqnarray}
where ${\widetilde\rho} = \mu/\sigma^2$. Under this measure, the identity $I_t$ is a martingale; however, the process $Z_t = S_0~\exp(\gamma X_t)~F(t)$ with $\gamma\neq 0$ is not a martingale for any function $F(t)$.

\section{Call Option Pricing Coefficients}\label{app.A}

{}The coefficients ${\widetilde c}_n$ for the call option price (\ref{gen-call}) in the model (\ref{tan}) are given by:
\begin{eqnarray}
 &&{\widetilde c}_0 = {a_0^2\over 4}\left(1-{k\over S_{mid}}\right)\left[2(L-x_*) + {\sin(\nu L) - \sin(\nu[2x_* - L])\over\nu}\right]-\nonumber\\
 &&-{a_0a_1\gamma k\over 2S_{mid}}\left[{{\cos([\lambda_1-\nu]L/2) - \cos([\lambda_1-\nu][x_*-L/2])}\over{\lambda_1-\nu}} +(\nu\rightarrow -\nu)\right]\\
 &&{\widetilde c}_1 = {a_0a_1\over 2}\left(1-{k\over S_{mid}}\right)\times\nonumber\\
 &&\times\left[{{\cos([\lambda_1-\nu]L/2) - \cos([\lambda_1-\nu][x_*-L/2])}\over{\lambda_1-\nu}} + (\nu\rightarrow -\nu)\right]-\nonumber\\
 &&-{a_1^2\gamma k\over 4S_{mid}}\left[2(L-x_*) - {\sin(\lambda_1 L) - \sin(\lambda_1[2x_* - L])\over\lambda_1}\right]\\
 &&{\widetilde c}_{n>1}={a_0a_n\over 2}\left(1-{k\over S_{mid}}\right)\times\nonumber\\
 &&\times\left[{{\sin\left({{[\lambda_n-\nu]L + \pi n}\over 2}\right) - \sin\left({{[\lambda_n-\nu](2x_*-L) + \pi n}\over 2}\right)}\over{\lambda_n-\nu}} + (\nu\rightarrow -\nu)\right]+\nonumber\\
 &&+{a_1a_n\gamma k\over 2S_{mid}}\left[{{\cos\left({{[\lambda_n-\lambda_1]L + \pi n}\over 2}\right) - \cos\left({{[\lambda_n-\lambda_1](2x_*-L) + \pi n}\over 2}\right)}\over{\lambda_n-\lambda_1}} - (\lambda_1\rightarrow -\lambda_1)\right]
\end{eqnarray}
where $a_n$ are given by (\ref{a.n.tan}), $\lambda_n$ are defined via (\ref{lambda.n.tan}), and $f(x_*)\equiv k$. These coefficients ${\widetilde c}_n$ reduce to those given by (\ref{c.0}), (\ref{c.1}) and (\ref{c.n}) in the $\nu\rightarrow 0$ limit.

\end{document}